\documentclass[aps,prl,superscriptaddress,twocolumn]{revtex4}

\usepackage{graphicx}
\usepackage{booktabs}

\begin{document}


\title{Local Structure and Spin Transition in Fe$_{2}$O$_{3}$ Hematite at High-Pressure}


\author{Andrea Sanson}
\email{andrea.sanson@unipd.it}
\affiliation{Department of Physics and Astronomy - University of
Padova, Padua (Italy)}
\author{Innokenty Kantor}
\affiliation{ESRF - European Synchrotron Radiation Facility,
Grenoble (France)}
\author{Valerio Cerantola}
\affiliation{ESRF - European Synchrotron Radiation Facility,
Grenoble (France)}
\author{Tetsuo Irifune}
\affiliation{Geodynamic Research Center - University of Ehime,
Matsuyama (Japan)}
\author{Alberto Carnera}
\affiliation{Department of Physics and Astronomy - University of
Padova, Padua (Italy)}
\author{Sakura Pascarelli}
\affiliation{ESRF - European Synchrotron Radiation Facility,
Grenoble (France)}


\date{\today}

\begin{abstract}
The pressure evolution of the local structure of Fe$_{2}$O$_{3}$
hematite has been determined for the first time by extended x-ray
absorption fine structure up to $\sim$79 GPa. The comparison to the
different high-pressure forms proposed in the literature suggests
that the orthorhombic structure with space group \emph{Aba2} is the
most probable. The crossover from Fe high-spin to low-spin states
with pressure increase has been monitored from the pre-edge region
of the Fe K-edge absorption spectra. The "simultaneous" comparison
with the local structural changes allows us to definitively conclude
that it is the electronic transition that drives the structural
transition and not viceversa.

\end{abstract}


\maketitle

The high-pressure behavior of hematite ($\alpha$-Fe$_{2}$O$_{3}$)
has raised much debate in the scientific community over the past
decades. At ambient conditions, hematite crystallizes in the
rhombohedral corundum-type structure, space group $R\overline{3}c$,
and is a wide-band antiferromagnetic insulator. By increasing
pressure at room temperature, the corundum structure of hematite is
progressively distorted and, above $\sim$50 GPa, a series of
physical changes occur
\cite{Pasternak99,Knittle86,Badro02,Rozenberg02,Liu03,Kozhevnikov07,Kunes09,Ghosh09,Wilson09,Wang10,Lin11}:
the unit cell volume drops down by about 10 \%, the crystal symmetry
changes completely, the electrical resistivity decreases drastically
due to the breakdown of the $d$-electron correlation (Mott
insulator-metal transition), the magnetic moments collapse
[transition of iron ions from high-spin (HS) to low-spin (LS) state]
and the long-range magnetic order disappears. Besides being
interesting from the viewpoint of solid-state physics, the phenomena
are also important in geophysics for modeling materials behavior in
deep Earth's mantle
\cite{Liebermann68,Kondo80,Yagi82,Shim02,Shim09}.

Despite many experimental and theoretical studies have been
conducted on this issue, several aspects still remain controversial
and unsolved. From the structural point of view, it was initially
proposed that the high-pressure (HP) form of Fe$_{2}$O$_{3}$ was a
GdFeO$_{3}$-type orthorhombic perovskite, containing two different
Fe sites with different coordination numbers and characterized by
unequal valence states, i.e., Fe$^{2+}$ and Fe$^{4+}$
\cite{Suzuki85,Olsen91,Syono84}. But a few years later, further
investigations established that the HP phase of Fe$_{2}$O$_{3}$ is a
non-magnetic metallic phase with a single Fe$^{3+}$ cation site, and
identified as the distorted Rh$_{2}$O$_{3}$-II structure
\cite{Pasternak99,Rozenberg02,Liu03}. However this conclusion cannot
be regarded as definitive because not based on x-ray
"single-crystal" diffraction data, that provides the most accurate
structural refinement, and moreover because certain aspects have
been challenged by Badro and co-workers \cite{Badro02}. In
confirmation of this, a recent synchrotron x-ray single-crystal
diffraction study proposes that, in the mixed state above 50 GPa,
Fe$_{2}$O$_{3}$ forms a novel monoclinic phase with space group
\emph{P2$_{1}$/n} and, above 67 GPa, compression triggers the
transition to a different HP phase with orthorhombic unit cell and
space group \emph{Aba2} \cite{Bykova13,Bykova15}. Another important
and controversial point refers to the nature of the phase-transition
at $\sim$50 GPa: is it the structural transition that drives the
electronic transition or viceversa? Some authors stated that the
structural transition precedes the change in the electronic
properties of Fe$_{2}$O$_{3}$ \cite{Badro02,Lin11}, while other
authors proposed the opposite scenario
\cite{Pasternak99,Rozenberg02,Kozhevnikov07,Kunes09,Ghosh09}.
Different theoretical approaches lead to different and controversial
results.

All these unsolved issues stimulate new high pressure studies of
hematite by means of other techniques. One of these is extended
x-ray absorption fine structure (EXAFS) spectroscopy that, thanks to
its selectivity to atomic species and insensitivity to long range
order, is a powerful tool for the study of local structure and
electronic properties of solids \cite{Stern88}. However, many
difficulties are associated with conducting EXAFS studies at
high-pressure, due to the strong absorption of the diamond anvils at
low x-ray energies (the Fe K-edge is at 7.1 KeV), and due to
intrinsic limitations to the measurable $k$-range for EXAFS because
of Bragg diffraction from the diamonds. In this study, Fe K-edge
energy dispersive EXAFS measurements have been conducted on
Fe$_{2}$O$_{3}$ under pressures up to $\sim$79 GPa (and then
decompressing up to $\sim$19 GPa) at the ID24 XAS beamline of the
European Synchrotron Radiation Facility (ESRF) in Grenoble
\cite{Pascarelli16}. The recent developments on this beamline,
together with the availability of nanodiamond anvils
\cite{Irifune03}, offer the possibility to reach a larger $k$-range
(up to about 10 {\AA}$^{-1}$ in the present study) at very high
pressures \cite{Baldini11,Torchio14}.
\begin{figure}
\includegraphics[width=0.33\textwidth]
{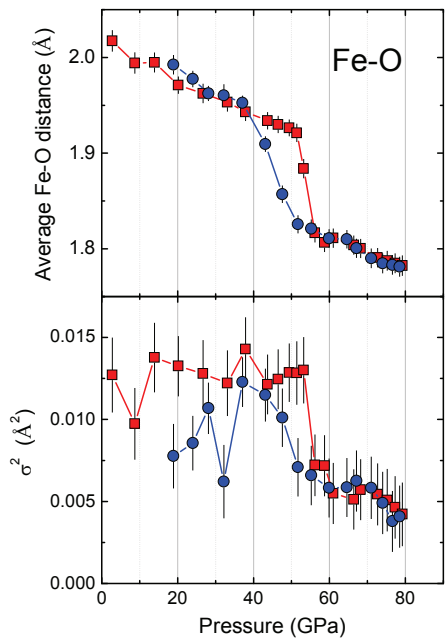} \caption{Pressure-dependence of the average Fe-O distance
(top panel) and of the variance $\sigma^{2}$ of corresponding Fe-O
distribution (bottom panel). Red squares and blue circles refer to
compression and decompression, respectively. The solid lines are
guide to the eyes.} \label{f.1}
\end{figure}

The experimental details and data analysis are provided in the
Supplemental Material \cite{SupplMaterial}. Due to the large number
of distances involved, the nearest-neighbor Fe-O distances were
assumed to follow one single-peak average distribution. The validity
of this assumption was tested in Ref. \cite{Sanson15} and is also
confirmed by the subsequent agreement with the crystallographic data
of hematite. The same assumption was made for the analysis of the
next-nearest-neighbor Fe-Fe/O distances. However, in the latter
case, the situation is much more complicated because the number (and
type) of atoms in the outer-shells is not well defined as in the
first-shell, where the Fe atoms are coordinated by 6 oxygens (within
a bonding distance of 2.3 {\AA}) in all the proposed HP structures
of Fe$_{2}$O$_{3}$ as in the simple corundum structure. Accordingly,
we will focus our attention on the nearest-neighbor Fe-O distances.
The results for the next-nearest-neighbor Fe-Fe/O distances are
reported and discussed in the Supplemental Material
\cite{SupplMaterial}.

The resulting average Fe-O distance as a function of pressure is
shown in the top panel of Fig. \ref{f.1}. The bottom panel of the
same figure shows instead the pressure dependence of variance
$\sigma^{2}$ of the corresponding average distance distribution. As
expected, it can be seen that the Fe-O distance progressively
decreases with increasing pressure (red squares), and, above
$\sim$50 GPa, we observe an abrupt decrease (of about 0.1 {\AA})
which reveals the HP phase transition of Fe$_{2}$O$_{3}$. Very
interesting is the behavior of $\sigma^{2}$ above the transition,
indicating a sharp decrease of the static disorder in the
nearest-neighbor Fe-O distances, consistent with a reduction in the
distortion of the FeO$_{6}$ octahedra. Finally, during
decompression, a hysteresis effect is observed across the phase
transition (blue circles in Fig. \ref{f.1}), although we do not have
direct evidence that the phase obtained upon decompression is
identical to the initial one. Indeed the $\sigma^{2}$ deviates
somewhat from that measured upon compression, and this could be
linked to different distortion of the FeO$_{6}$ octahedra.
\begin{figure}
\includegraphics[width=0.28\textwidth]
{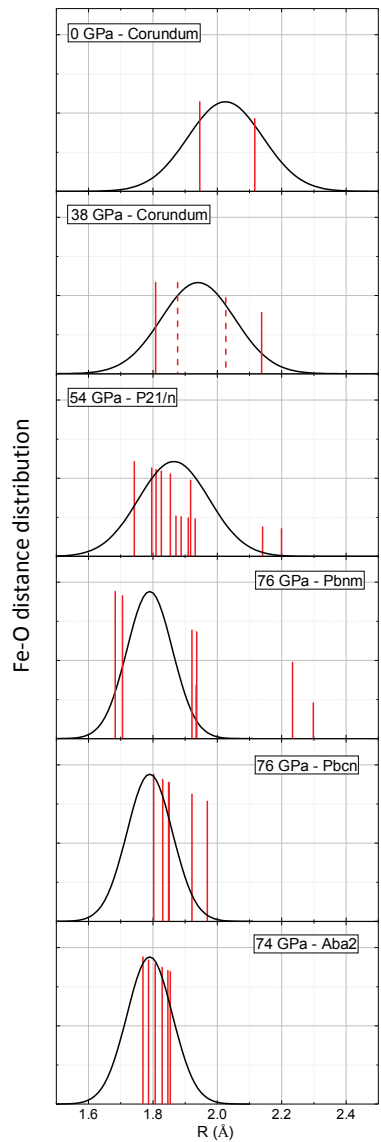} \caption{Fe-O average distributions determined by EXAFS
(black-solid lines). The red bars indicate the Fe-O distances
according to the HP structures proposed in literature. The
red-dashed vertical bars in the second panel from the top are the
Fe-O distances of hematite by assuming unchanged the atomic
positions in the unit cell.} \label{f.2}
\end{figure}
\begin{table*}
\caption{Average Fe-O parameters expected for the different HP
structures of Fe$_{2}$O$_{3}$ (left side) and their comparison with
the experimental EXAFS results (right side). In order to do a more
accurate comparison, the structural parameters were calculated by
weighting the respective distances according to their EXAFS
scattering-amplitudes, while the dynamic contribution to
$\sigma^{2}$ was approximated to those of hematite at ambient
conditions (see text).} \label{TabI}
\begin{ruledtabular}
\begin{tabular}{lllll}
\toprule
Crystal Structure &  \multicolumn{2}{c}{Expected} & \multicolumn{2}{c}{EXAFS}\\
\midrule & Fe--O [{\AA}]   &   $\sigma^{2}_{Fe-O}$ [{\AA}$^{2}$] &
Fe--O [{\AA}] &  $\sigma^{2}_{Fe-O}$ [{\AA}$^{2}$]
 \tabularnewline \hline
Corundum (0 GPa) & 2.02 & 0.013 &2.02$\pm$0.01  &0.013$\pm0.002$\\
Corundum A (38GPa)$^{a}$ &  1.94 &0.032 &1.94$\pm$0.01 &0.013$\pm$0.002\\
Corundum B (38GPa) &  1.94 &   0.012 &1.94$\pm$0.01 &0.013$\pm$0.002 \\
P21/n (54 GPa)$^{b}$ & 1.87 & 0.017 &1.87$\pm$0.01  &0.012$\pm$0.002  \\
Pbnm (76 GPa)$^{a,*}$ & 1.81/1.88 & 0.020/0.046 &1.79$\pm$0.01  &0.005$\pm$0.002  \\
Pbcn (76 GPa)$^{a}$ & 1.87 & 0.009 &1.79$\pm$0.01  &0.005$\pm$0.002  \\
Aba2 (74 GPa)$^{b}$ & 1.81 & 0.007 &1.79$\pm$0.01  &0.005$\pm$0.002 \\
\end{tabular}
a) crystal cell from Ref. \cite{Rozenberg02}, b) from Refs.
\cite{Bykova13,Bykova15}.\\
$^{*}$the two set of values were obtained by neglecting or including
the Fe-O distances at $\sim$2.3 {\AA}, respectively.
\end{ruledtabular}
\end{table*}

Fig. \ref{f.2} shows the average Fe-O distance distributions at the
pressures of interest for our next discussion, roughly approximated
to Gaussian distributions in accordance to the data of Fig.
\ref{f.1}. The vertical bars indicate the Fe-O distances according
to the HP structures proposed in literature
\cite{Suzuki85,Olsen91,Syono84,Pasternak99,Rozenberg02,Liu03,Bykova13,Bykova15}.
Their heights were scaled according to the EXAFS amplitude
calculated by the FEFF code \cite{FEFF,FEFF2} at 0 K (atoms frozen
in their equilibrium positions) and arbitrarily normalized to the
height of the Fe-O distributions.

The local structure results of Fig. \ref{f.2}, summarized in Tab.
\ref{TabI}, can help to shed light on the controversial HP phase of
Fe$_{2}$O$_{3}$. Note that the EXAFS $\sigma^{2}$ reported in the
last column of Tab. \ref{TabI}, is the sum of a static contribution
$\sigma^{2}_{st}$ due to the presence of Fe-O distances of different
lengths, and a dynamic contribution $\sigma^{2}_{din}$ due to
thermal disorder. For the Fe-O nearest-neighbors of hematite at
ambient conditions, $\sigma^{2}_{st}$ is about 0.007 {\AA}$^{2}$,
while the average value of $\sigma^{2}_{din}$, determined by
temperature-dependent EXAFS measurements and molecular dynamics
calculations \cite{Sanson14}, is about 0.006 {\AA}$^{2}$. In the
determination of the expected $\sigma^{2}$ of the different HP
structures, listed in the third column of Tab. \ref{TabI}, we
approximated the dynamic contribution $\sigma^{2}_{din}$ to that of
hematite at ambient conditions.

As first, we consider the HP structure before the transition
proposed by Rozenberg and co-workers \cite{Rozenberg02}. According
to their refined structural parameters, the resulting average Fe-O
distance is in very good agreement with that obtained in the present
EXAFS study (2nd panel from the top of Fig. \ref{f.2}). However,
Rozenberg \emph{et al.} found that pressure induces a progressive
distortion of the FeO$_{6}$ octahedron, in which the distance gap
between short and long Fe-O nearest-neighbors progressively
increases up to about 0.4 {\AA}, resulting in a progressive increase
of the static disorder $\sigma^{2}_{st}$ up to $\sim$0.04
{\AA}$^{2}$. This is in sharp contrast with our EXAFS results shown
in the bottom panel of Fig. \ref{f.1}, where below the
phase-transition $\sigma^{2}$ is constant and $\sim$0.013
{\AA}$^{2}$. Accordingly, no further distortion of the FeO$_{6}$
octahedra is observed with increasing pressure with respect to the
ambient conditions. To show this, the average Fe-O structural
parameters were calculated from Rozenberg's refinement, for example
at 38 GPa (Tab. \ref{TabI} - Corundum A), and the same was done
using the lattice parameters of Rozenberg \emph{et al.} but leaving
unchanged the atomic positions in the unit cell (Tab. \ref{TabI} -
Corundum B). It can be seen that the agreement with the experimental
EXAFS data, specifically $\sigma^{2}$, is much better for the
latter.

We now consider the HP structures of Fe$_{2}$O$_{3}$ above the phase
transition. For the novel monoclinic phase (space group
\emph{P21/n}) proposed by Bykova \emph{et al.}
\cite{Bykova13,Bykova15} at 54 GPa, only the expected average Fe-O
distance is in agreement with that of EXAFS (Tab. \ref{TabI}, 4th
row); in contrast, the value of $\sigma^{2}$ indicates a small, but
not negligible, Fe-O distortion and therefore we cannot validate
this HP structure. More enlightening is the comparison of the HP
structures above 70 GPa. The local structural parameters of the
GdFeO$_{3}$-type perovskite structure, space group \emph{Pbnm}, are
completely at odds with the EXAFS results (Tab. \ref{TabI}, 5th
row), independently on whether the Fe-O distances at 2.3 {\AA} are
included or not. In particular, the static disorder of the
nearest-neighbor Fe-O distribution is too large compared to that
measured by EXAFS, therefore, in agreement with previous studies
\cite{Pasternak99,Rozenberg02,Liu03,Shim02}, we can rule out the
GdFeO$_{3}$ form as HP structure of Fe$_{2}$O$_{3}$. However, we
come to the same conclusion also for the distorted
Rh$_{2}$O$_{3}$-II structure, space group \emph{Pbcn}, which is
currently the most accepted HP structure for Fe$_{2}$O$_{3}$
\cite{Pasternak99,Rozenberg02,Liu03}. Indeed, from Tab. \ref{TabI} -
6th row, the structural parameters of this form show a significant
discrepancy with the EXAFS results. On the contrary, the best
agreement seems to be found for the orthorhombic structure with
space group \emph{Aba2} (Tab. \ref{TabI}, 7th row), the HP structure
very recently proposed by Bykova and co-workers \cite{Bykova15} on
the basis of synchrotron x-ray single-crystal diffraction.

\begin{figure}
\includegraphics[width=0.35\textwidth]
{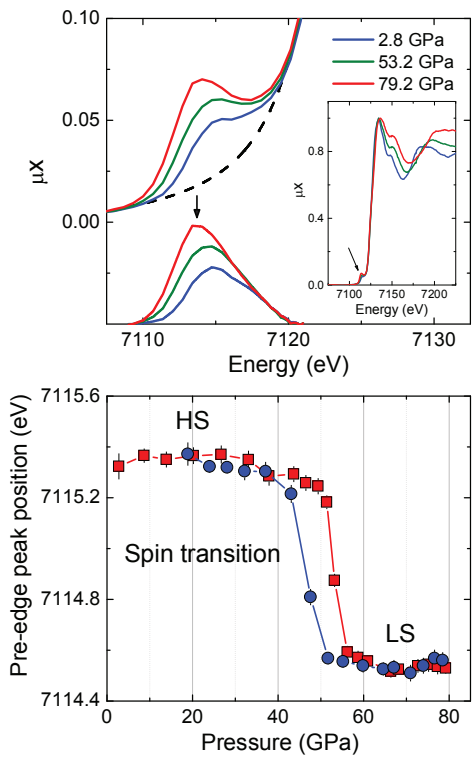} \caption{Top panel: pre-edge peak of the Fe K-edge
absorption spectra at selected pressures and corresponding
background subtraction. The inset shows the whole absorption
spectra. Bottom panel: Fe$^{3+}$ high-spin/low-spin crossover
monitored by the average pre-edge peak position as a function of
pressure, during compression (red squares) and decompression (blue
circles).} \label{f.3}
\end{figure}

We now address the controversial issue of the nature of the
phase-transition
\cite{Badro02,Lin11,Pasternak99,Rozenberg02,Kozhevnikov07,Kunes09,Ghosh09},
i.e., do the electronic properties of Fe$_{2}$O$_{3}$ change only
after the structural transition with a decrease in volume and a
change in the lattice symmetry, or viceversa? In this regard, the
evolution of the Fe 3d electronic structure vs pressure can be
investigated from the pre-edge region of the Fe K-edge absorption
spectrum (top panel of Fig. \ref{f.3}), since this pre-edge feature
can be assigned to transitions to the t2g and eg components of the
3d band \cite{Wang10,Caliebe98} and, therefore, is directly
connected to the population of the HS and LS states. After
normalization and background subtraction of absorption spectra (top
panel of Fig. \ref{f.3}), the average position of the pre-edge peak
was determined and plotted as a function of pressure (bottom panel
of Fig. \ref{f.3}): in this way we monitor the crossover from HS to
LS states with pressure increase and viceversa.
\begin{figure}
\includegraphics[width=0.33\textwidth]
{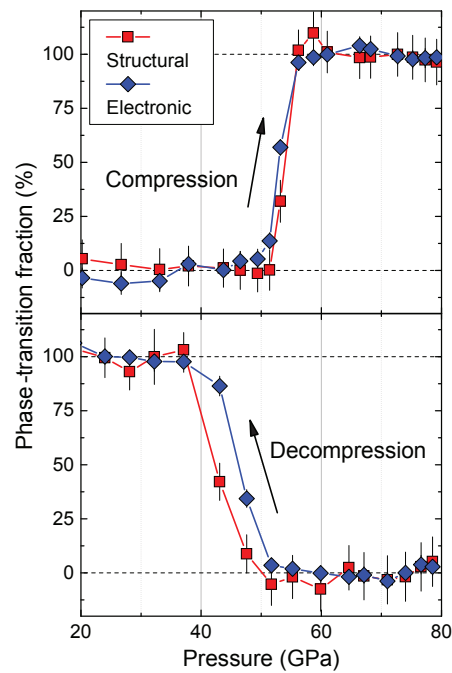} \caption{Phase-transition fraction (in percent) during
compression (top panel) and decompression (bottom panel).
Red-squares refer to the structural transition, blue-diamonds to the
electronic transition. In the both cases, the electronic transition
precedes the structural transition.} \label{f.4}
\end{figure}

The "simultaneous" measurement of the structural transition (Fig.
\ref{f.1}) and of the electronic transition (Fig. \ref{f.3}) is
fundamental to overcome the uncertainty due to different hydrostatic
conditions and different measured volume, and allows addressing
questions related to the interplay between structural and
electronic/magnetic degrees of freedom in Fe$_{2}$O$_{3}$, as
previously demonstrated for pure Fe \cite{Mathon04}. In Fig.
\ref{f.4} we show the evolution of the phase-transition fraction,
both structural and electronic, during compression (top panel) and
decompression (bottom panel). During compression, in the mixed HS/LS
state at about 53 GPa, the electronic transition is $\sim$60 \%
completed (blue diamonds in the top panel of Fig. \ref{f.4}), while
the structural transition (red squares) is only $\sim$30 \%
completed. As a result, we can deduce that the HP structural
transition occurs only after the electronic transition to the
low-spin phase. Further confirmation of this finding is given by the
data collected during decompression. At about 48 and 43 GPa, the LS
to HS transition is completed at $\sim$35 \% and $\sim$85 \%,
respectively (blue diamonds in the bottom panel of Fig. \ref{f.4}),
while the structural transition (red squares) is only completed at
$\sim$10 \% and $\sim$40 \%, respectively. This shows again that the
electronic transition actually precedes the structural transition,
and leads to the following description of the HP transition of
Fe$_{2}$O$_{3}$: \emph{i}) volume and bond distances decrease with
pressure until a "volume threshold" value is reached at $\sim$50
GPa, in which the low-spin phase is more stable as predicted by some
theoretical calculations \cite{Rollmann04,Kunes09,Ghosh09}.
The spin crossover transition from HS to LS states is thus activated
\emph{ii}) the full HS to LS transition occurs between $\sim$50 and
55 GPa. In this pressure interval, the sample enters into a
metastable phase characterized by Fe$^{3+}$ ions in the mixed HS/LS
state, in which the crystal structure is still (at least initially)
unchanged. Further increase in pressure triggers the Fe$^{3+}$ ions
into the stable LS phase, thus causing the structural transition and
volume collapse (LS Fe atomic radius becomes $\sim$0.1 {\AA} shorter
than that of HS Fe \cite{Shannon76}) as a consequence of emptying of
the anti-bonding bands in the LS phase and corresponding
strengthening of the Fe-O bonds \cite{Kunes09} \emph{iii}) above
$\sim$55 GPa the Fe$^{3+}$ are all (or almost) LS, and bond
distances and volume continue to decrease as a mere effect of
pressure increase.

Summarizing, in this Letter we have studied the local structure of
Fe$_{2}$O$_{3}$ hematite under high pressure. Below the phase
transition, no increasing FeO$_{6}$ octahedra distortion is observed
as pressure is applied, in contrast to Rozenberg et al.
\cite{Rozenberg02}. More importantly, an abrupt decrease in the
nearest-neighbor Fe-O distance is observed at $\sim$50 GPa.
Concomitantly, we observe a peculiar decrease of the
nearest-neighbor Fe-O static disorder, indicating a reduction in the
FeO$_{6}$ distortion. The present EXAFS results represent an
excellent test-bench for proposed or new HP forms of
Fe$_{2}$O$_{3}$. Comparison to the different HP phases proposed in
the literature rules out the GdFeO$_{3}$-type orthorhombic
perovskite form as well as the most accepted distorted
Rh$_{2}$O$_{3}$-II structure, and rather suggests that the
orthorhombic structure with space group \emph{Aba2} is the most
appropriate among those reported in literature.

Finally, the pressure-induced Fe$^{3+}$ high-spin to low-spin
transition has been monitored from the pre-edge peak of the Fe
K-edge absorption spectra. The simultaneous comparison with the
pressure evolution of the local structural transition determined by
EXAFS allows us to conclude that it is the electronic transition
that drives the structural transition and not viceversa, thus
definitively solving the longstanding controversy on the nature of
the phase-transition. The details of the dynamics of this phase
transition, and in particular, the nature of the observed metastable
phase, call for further theoretical and experimental investigations.

We acknowledge the European Synchrotron Radiation Facility (ESRF)
for provision of synchrotron radiation, as well as E. Bykova, L.
Dubrovinsky and O. Mathon for useful discussions. This work has been
partially supported by the ESRF project No. HE-3766.

%
%

\

\vfill\eject

\end{document}